\documentclass[review,number,sort&compress]{elsarticle}
\usepackage{lineno}

\usepackage{amssymb,amsmath,latexsym}                        
\usepackage{graphicx}
\usepackage{cite}
\usepackage{xcolor}
\usepackage{float}

\usepackage{amsthm}
\usepackage{multicol}

\begin{document}

\begin{frontmatter}
\title{Mathematical Properties of \\Numerical Inversion for Jet Calibrations}
\author[a,b]{Aviv Cukierman}
\author[a,b,c]{Benjamin Nachman}
\address[a]{Physics Department, Stanford University, Stanford, CA, 94305, USA}
\address[b]{SLAC National Accelerator Laboratory, Stanford University, Menlo Park, CA 94025, USA}
\address[c]{Physics Division, Lawrence Berkeley National Laboratory, Berkeley, CA 94704, USA}


\begin{abstract}
Numerical inversion is a general detector calibration technique that is independent of the underlying spectrum.  This procedure is formalized and important statistical properties are presented, using high energy jets at the Large Hadron Collider as an example setting.  In particular, numerical inversion is inherently biased and common approximations to the calibrated jet energy tend to over-estimate the resolution.  Analytic approximations to the closure and calibrated resolutions are demonstrated to effectively predict the full forms under realistic conditions.  Finally, extensions of numerical inversion are presented which can reduce the inherent biases.  These methods will be increasingly important to consider with degraded resolution at low jet energies due to a much higher instantaneous luminosity in the near future.
\end{abstract}

\end{frontmatter}

\section{Introduction}
\label{sec:intro}

\noindent At a proton-proton collider like the Large Hadron Collider (LHC), quarks and gluons are produced copiously.  These partons fragment to produce collimated streams of colorless particles that leave their energy in the calorimeters of the ATLAS and CMS detectors\footnote{Jets have been calibrated in previous experiments, such as the Tevatron CDF~\citep{Bhatti:2005ai} and D0~\citep{Abazov:2013hda} experiments, but the methods were significantly different and so this note focuses on the general purpose LHC experiments.}.  The energy depositions are organized using jet clustering algorithms to stand as experimental proxies for the initiating quarks and gluons.  The most widely used clustering scheme in ATLAS and CMS is the anti-$k_t$ algorithm~\citep{Cacciari:2008gp} with radius parameter $R=0.4$.  Even though the inputs to jet clustering (topological clusters for ATLAS~\citep{topo1,topo2} and particle flow objects for CMS~\citep{pflow1,pflow2}) are themselves calibrated, the average reconstructed jet energy is not the same as the true jet energy, because of various detector effects.  To account for this, calibrations are applied to each reconstructed jet.  

\section{Numerical Inversion}
\label{sec:numinversion}

The jet calibration procedures of ATLAS~\citep{Aad:2011he} and CMS~\citep{Chatrchyan:2011ds,Khachatryan:2016kdb} involve several steps to correct for multiple nearly simultaneous $pp$ collisions (pileup), the non-linear detector response, the $\eta$-dependence of the jet response, flavor-dependence of the jet response, and residual data/simulation differences in the jet response.  The simulation-based corrections to correct for the calorimeter non-linearities in transverse energy $E_\text{T}$ and pseudorapidity $\eta$ are accounted for using {\it numerical inversion}.

The purpose of this note is to formally document numerical inversion and describe (with proof) some of its properties.   In what follows, $X$ will be a random variable representing the particle-jet $E_\text{T}$ and $Y$ will be a random variable representing the reconstructed jet $E_\text{T}$.  Define\footnote{Capital letters represent random variables and lower case letters represent realizations of those random variables, i.e. $X=x$ means the random variable $X$ takes on the (non-random) value $x$.}

\begin{align}
\label{eq:fedef}
f_\text{me}(x)&=\mathbb{E}[Y|X=x]\\\label{eq:fedef2}
R_\text{me}(x) &= \mathbb{E}\left[\frac{Y}{x}\middle| X=x\right] = \frac{f_\text{me}(x)}{x}. 
\end{align}

Where the subscript indicates that we are taking the mean of the stated distribution and `$\mathbb{E}$' stands for {\it expected value} ($=$ average). In practice, sometimes the core of the distribution of $Y|X=x$ is fit with a Gaussian and so the effective measure of central tendency is the mode of the distribution.  Therefore in analogy to Equations~\ref{eq:fedef} and~\ref{eq:fedef2}, we define
\begin{align}
f_\text{mo}(x)&=\text{mode}[Y|X=x]\\
R_\text{mo}(x) &= \text{mode}\left[\frac{Y}{x}\middle| X=x\right] = \frac{f_\text{mo}(x)}{x}. 
\end{align}
We will often drop the subscript of $f$ and $R$ for brevity in the text, when it is clear which definition we are referring to. If not specified, $f$ and $R$ will refer to a definition using a generic definition of central tendency.  For all sensible notions of central tendency, we still have that $R(x) = \frac{f(x)}{x}$.

We will often think of $Y|X=x\sim \mathcal{N}(f(x),\sigma(x))$, where this notation means `$Y$ given $X=x$ is normally distributed with mean $f(x)$ and standard deviation $\sigma(x)$'; however, in this note, we will remain general unless stated otherwise.  The function $R(x)$ is called the {\it response function}.  Formally, numerical inversion is the following procedure:

\begin{enumerate}
\item Compute $f(x)$, $R(x)$.  
\item Let $\tilde{R}(y) = R(f^{-1}(y))$.
\item Apply a jet-by-jet correction: $Y\mapsto Y/\tilde{R}(Y)$.
\end{enumerate}

\noindent The intuition for step 2 is that for a given value $y$ drawn from the distribution $Y|X=x$, $f^{-1}(y)$ is an estimate for $x$ and then $R(f^{-1}(y))$ is an estimate for the response at the value of $x$ that gives rise to $Y$. Let $p(x)$ be the prior probability density function of $E_\text{T}$. Then we note that we do not want to use $\mathbb{E}[X|Y]$ instead of $f^{-1}(Y)$ because the former depends on $p(x)$, whereas $f$ (and thus $f^{-1}$) does not depend on $p(x)$, by construction.  

\newpage
We can see now our first result, which will be useful for the rest of this note:
\vspace{6mm}

\noindent{\it The correction derived from numerical inversion is $Y \mapsto Z = f^{-1}(Y)$.}

\noindent{\bf Proof.}
\begin{align}
\tilde{R}(Y) &= R(f^{-1}(Y))\nonumber\\
&= \frac{f(f^{-1}(Y))}{f^{-1}(Y)}\nonumber\\
&= \frac{Y}{f^{-1}(Y)}\nonumber\\
\rightarrow Z &= \frac{Y}{\tilde{R}(Y)}\nonumber\\
&= f^{-1}(Y) \hspace{1 cm} \Box
\end{align}

\subsection{Closure}
\label{sec:introclosure}
One important property of numerical inversion is the concept of {\it closure}, which quantifies whether the new distribution $f^{-1}(Y|X=x)$ obtained after numerical inversion is centered at $x$, using the same notion of central tendency as in the definition of $f$.  In particular, define the closure as

\begin{align}
C_\text{me}(x) \equiv \mathbb{E}\left[\frac{Z}{x}\middle| X=x\right] = \mathbb{E}\left[\frac{f^{-1}(Y)}{x}\middle| X=x\right]
\label{eqn:closure},
\end{align}
and $C_\text{mo}$ is defined in an analogous way.  The symbol $C$ will denote the closure for a generic notion of central tendency.  We say that numerical inversion has {\it achieved closure} or simply {\it closes} if, for all $x$,
\begin{align}
C = 1.
\end{align}

\subsection{Assumptions and Definitions}
\label{sec:assumptions}

The general results presented in the following sections are based on three assumptions listed below.  These requirements should be satisfied by real detectors using calorimeters and trackers to reconstruct jets, given that the detector-level reconstruction is of sufficiently high quality.

\begin{enumerate}
\item $f^{-1}(y)$ exists for all $y$ in the support of $Y$, and $f^{-1}$ is single-valued.  These may seem like obvious statements, but are not vacuous, even for a real detector.  For example, pileup corrections can result in non-zero probability that $Y<0$, so the function $f$ must be computed for all possible values of $Y$, even if the transverse energy is negative.  At the high-luminosity LHC (HL-LHC), the level of pileup will be so high that the jet energy resolution may be effectively infinite at low transverse energies (no correlation between particle-level and detector-level jet energy).  In that case, $f^{-1}$ may not be single valued and numerical inversion cannot be strictly applied as described in Sec.~\ref{sec:numinversion}.
\item $f(x)$ is monotonically increasing: $f'(x)>0$ for all $x$.  This condition should trivially hold for any reasonable detector: detector-level jets resulting from particle-level jets with a higher $E_\text{T}$ should on average have a higher $E_\text{T}$ than those originating from a lower $E_\text{T}$ particle-level jet.  Note that this is only true for a fixed $\eta$.  Detector technologies depend significantly on $\eta$ and therefore the $\eta$-dependence of $f$ (for a fixed $x$) need not be monotonic. We note also that Assumption 1 implies that $f'(x)\ge 0$ or $f'(x) \le 0$ for all $x$; so Assumption 2 is equivalent to the additional assumptions that $f'(x)\ne 0$ for any $x$, and that $f'(x)>0$ (as opposed to $f'(x)<0$).
\item $f$ is twice-differentiable. The first derivative of $f$ has already been assumed to exist in Assumption 2, and the second derivative will also be required to exist for some of the later results. In practice we expect $f$ to be differentiable out to any desired order.
\end{enumerate}

We note that as long as the above three assumptions hold, the theorems stated in the remainder of this paper are valid. In particular, this implies that $x$ could be any calibrated quantity that satisfies the above constraints, e.g. the jet transverse momentum $p_\text{T}$ or the jet mass $m$. We focus on the case of calibrating the $E_\text{T}$ for sake of concreteness.

We have separated the results in this paper into ``Proofs'' and ``Derivations''. The ``Proofs'' require only the three assumptions stated above, and in particular do not assume anything about the shape of the underlying distributions, e.g. that the distributions $Y|X=x$ are Gaussian or approximately Gaussian. The ``Derivations'' are useful approximations that apply in the toy model described in~\ref{sec:toy_model}; we expect them to apply in a wide variety of cases relevant to LHC jet physics. In particular, we expect these approximations to hold in cases with properties similar to the toy model presented here - e.g., good approximation of $f$ by its truncated Taylor series about each point and approximately Gaussian underlying distributions of $Y|X=x$.\footnote{Note that we do not require that $Y|X=x$ is exactly Gaussian, only that it is approximately Gaussian, which is true for a wide range of energies and jet reconstruction algorithms at ATLAS and CMS. In particular, there are non-negligible (but still often small) asymmetries at low and high $E_\text{T}$ at ATLAS and CMS~\citep{Aad:2011he,Chatrchyan:2011ds,Khachatryan:2016kdb}. In any case, even if $Y|X=x$ {\it is} Gaussian, $Z|X=x$ is in general {\it not} Gaussian, for non-linear response functions; see~\ref{sec:lemma}.}

Finally, in the rest of this paper, we write $\rho_{Y|X}(y|x)$ to represent the probability distribution of $Y$ given $X=x$, and $\rho_{Z|X}(z|x)$ to be the probability distribution of $Z$ given $X=x$. A standard fact about the probability distribution from changing variables is that

\begin{align}
\rho_{Z|X}(z|x) = f'(z)\rho_{Y|X}(f(z)|x).
\label{eqn:newdist}
\end{align}

\noindent To ease the notation, we will often use $\rho_Y(y)$ and $\rho_Z(z)$ interchangeably with $\rho_{Y|X}(y|x)$ and $\rho_{Z|X}(z|x)$, respectively, when it is clear (as is usually the case) that we are conditioning on some true value $x$.\footnote{In practice it is necessary to condition on a small range of $X$, e.g. $X\in[x,(1+\epsilon)x]$. If $\epsilon$ is large then there can be complications from the changing of $f(x)$ over the specified range and from the shape of the prior distribution of $X$ over the specified range.  These challenges can be solved by generating large enough Monte Carlo datasets.  We therefore assume that $\epsilon \ll 1$ and consider complications from finite $\epsilon$ beyond the scope of this paper.}

\section{Results}

In the subsequent sections, we will derive properties about the closure $C$ for three different definitions of the central tendancy: mean (Sec.~\ref{sec:mean}), mode (Sec.~\ref{sec:mode}), and median (Sec.~\ref{sec:median}).

\subsection{Mean}
\label{sec:mean}

In the following section only, for brevity, we will let $f$ be $f_\text{me}$ and $C$ be $C_\text{me}$.
\subsubsection{Closure}
\label{sec:meanclosuresection}

We can write the closure (Eq. ~\ref{eqn:closure}) as

\begin{align}
C = \mathbb{E}\left[\frac{Z}{x}\middle| X=x\right] &=\frac{1}{x} \int dy \rho_{Y|X}(y|x) f^{-1}(y).
\label{eqn:closuredef}
\end{align}

\noindent We find that for many functions $f$, numerical inversion does not close. This is summarized in the following result:

\vspace{5mm}

\noindent {\it Let the notion of central tendency be the mean. If $f$ is linear, then numerical inversion closes. If $f$ is not linear, then numerical inversion does not necessarily close.}

\vspace{5mm}

\noindent {\bf Proof.}
Let $f$ be linear, $f(x) = a(x+b)$. Then\footnote{We have $a>0$ from the assumption that $f'(x)>0$.} $f^{-1}(y) = \frac{y}{a}-b$. We can see that we necessarily have closure as Eq.~\ref{eqn:closuredef} can be written
\begin{align}
C &=\frac{1}{x} \int dy \rho_{Y|X}(y|x) \left(\frac{y}{a}-b\right)\nonumber\\
&=\frac{1}{x} \left(\frac{1}{a}\mathbb{E}\left[Y\middle| X=x\right]-b\right)\nonumber\\
&=\frac{1}{x} \left(\frac{1}{a}f(x)-b\right)\nonumber\\
&=1.
\label{eqn:closure_linear_proof}
\end{align}

Now let $f$ be nonlinear, and so therefore $f^{-1}$ is also nonlinear. We note that the statement being proved is that $f$ does not necessarily close in this case; not that $f$ necessarily does not close. Thus, it is sufficient to find a counterexample that does not close in order to demonstrate this statement. Let $f(x) = \left(\frac{x}{c}\right)^{\frac{1}{3}}$ with $c\ne 0$, so that $f^{-1}(y) = cy^3$, which is a simple non-linear monotonic function. We will also need to specify some higher moments of the distribution $\rho_{Y|X}$. With the standard definitions of the variance and skew, respectively:
\begin{align}
\sigma(x)^2&\equiv
\mathbb{E}\left[\left(Y-\mathbb{E}\left[Y\right]\right)^2\middle| X=x\right]\\
\sigma(x)^3\gamma_1(x) &\equiv \mathbb{E}\left[\left(Y-\mathbb{E}\left[Y\right]\right)^3\middle| X=x\right].
\end{align}
We specify the weak conditions that $\sigma(x) >0$ (which is always true as long as $\rho_{Y|X}$ is not a delta function), and that $\gamma_1(x)=0$ (which is true if $\rho_{Y|X}$ is symmetric).  Then, the closure (Eq.~\ref{eqn:closuredef}) can be written
\begin{align}
C &=\frac{1}{x} \int dy \rho_{Y|X}(y|x) \left(cy^3\right)\nonumber\\
&=\frac{c}{x} \left(\mathbb{E}\left[Y^3\middle| X=x\right]\right).
\end{align}
With $\gamma_1(x)=0$, we have that
\begin{align}
\mathbb{E}\left[Y^3\middle| X=x\right] &= 3\sigma(x)^2\mathbb{E}\left[Y\middle| X=x\right] + \mathbb{E}\left[Y\middle| X=x\right]^3\nonumber\\
&=3\sigma(x)^2f(x)+f(x)^3\nonumber\\
&=3\sigma(x)^2\left(\frac{x}{c}\right)^{\frac{1}{3}}+\frac{x}{c}.
\end{align}
Then we see we do not have closure, as
\begin{align}
C &=\frac{c}{x} \left(\mathbb{E}\left[Y^3\middle| X=x\right]\right)\nonumber\\
&=\frac{c}{x} \left(3\sigma(x)^2\left(\frac{x}{c}\right)^{\frac{1}{3}}+\frac{x}{c}\right)\nonumber\\
&= 1 + 3\sigma(x)^2\left(\frac{x}{c}\right)^{-\frac{2}{3}}\nonumber\\
&>1. \hspace{1 cm}\Box
\end{align}
Although the counterexample provided here only applies to a specific choice of $f(x)$ and $\rho_{Y|X}(y|x)$, we have reason to believe that closure is not achieved for non-linear $f$ in the vast majority of cases, as can be seen in more detail in~\ref{sec:mean_nonclosure}. In addition, we can Taylor expand the closure $C$ to derive an equation for the first non-closure term:
\begin{align}
C \approx 1-\frac{1}{2}\frac{f''(x)}{f'(x)^3}\frac{\sigma(x)^2}{x},
\label{eqn:closureseries_text}
\end{align}
the derivation of which can be found in~\ref{sec:mean_nonclosure}.

\begin{figure}[]
\begin{center}
\includegraphics[width=0.9\textwidth]{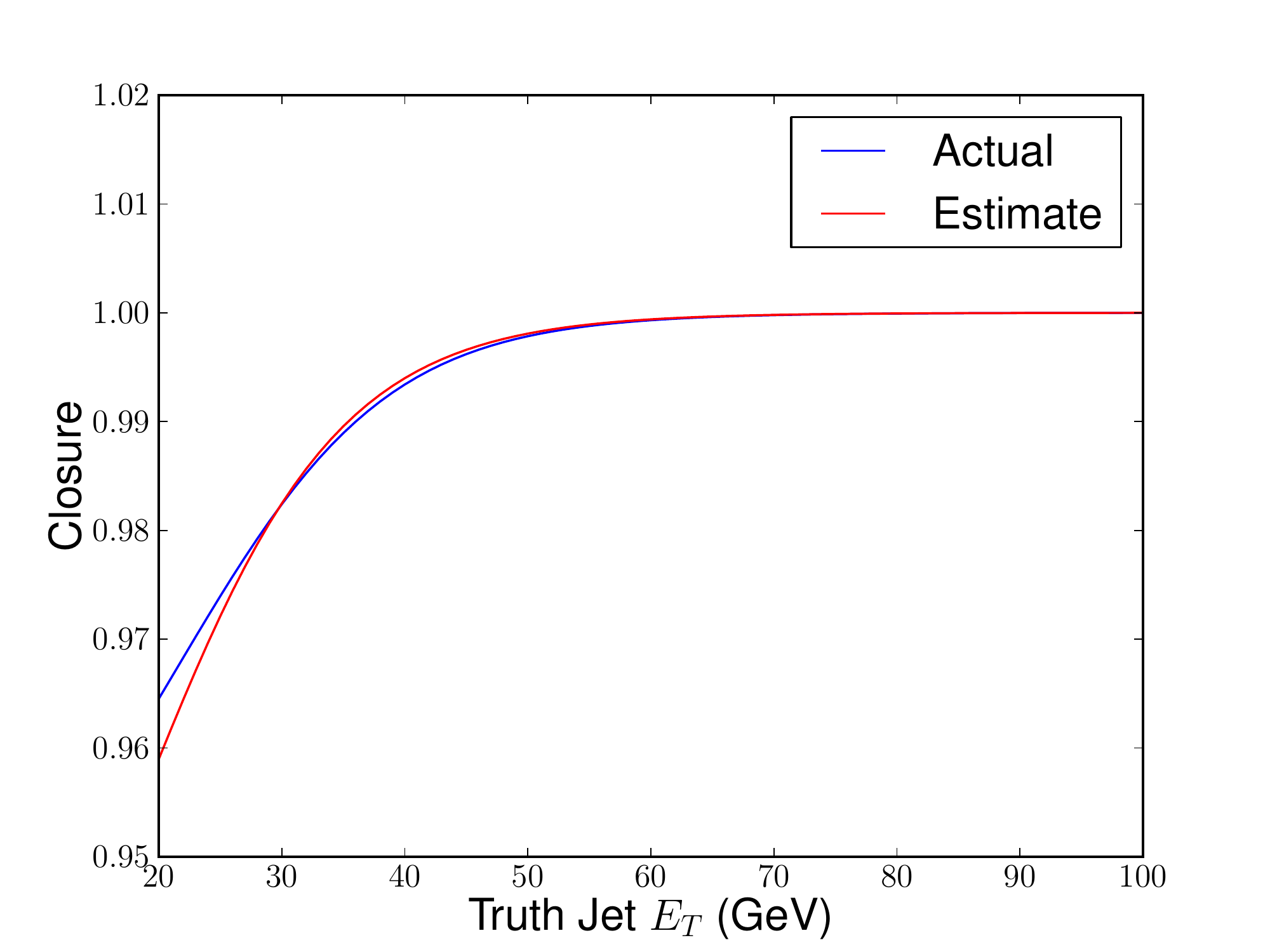}
\end{center}
\caption{The closure of numerical inversion when using the mean to calibrate, using a toy model similar to conditions in ATLAS or CMS. In blue, the exact calculated closure. In red, the estimate of the closure using the first term of the higher-order correction given in Eq.~\ref{eqn:closureseries_text}. For details of the model, see~\ref{sec:toy_model}.}
\label{fig:mean_closure}
\end{figure}

Figure~\ref{fig:mean_closure} shows the inherent non-closure in numerical inversion for a toy calculation using a response function $R(x)$ that is typical for ATLAS or CMS, and the first term of the higher-order correction (Eq.~\ref{eqn:closureseries_text}).

\subsubsection{Calibrated Resolution}
We often care about how well we have resolved the transverse energy of the jets, which we quantify by examining the width of the calibrated resolution $Z$.

\begin{figure}
\begin{center}
\includegraphics[width=0.9\textwidth]{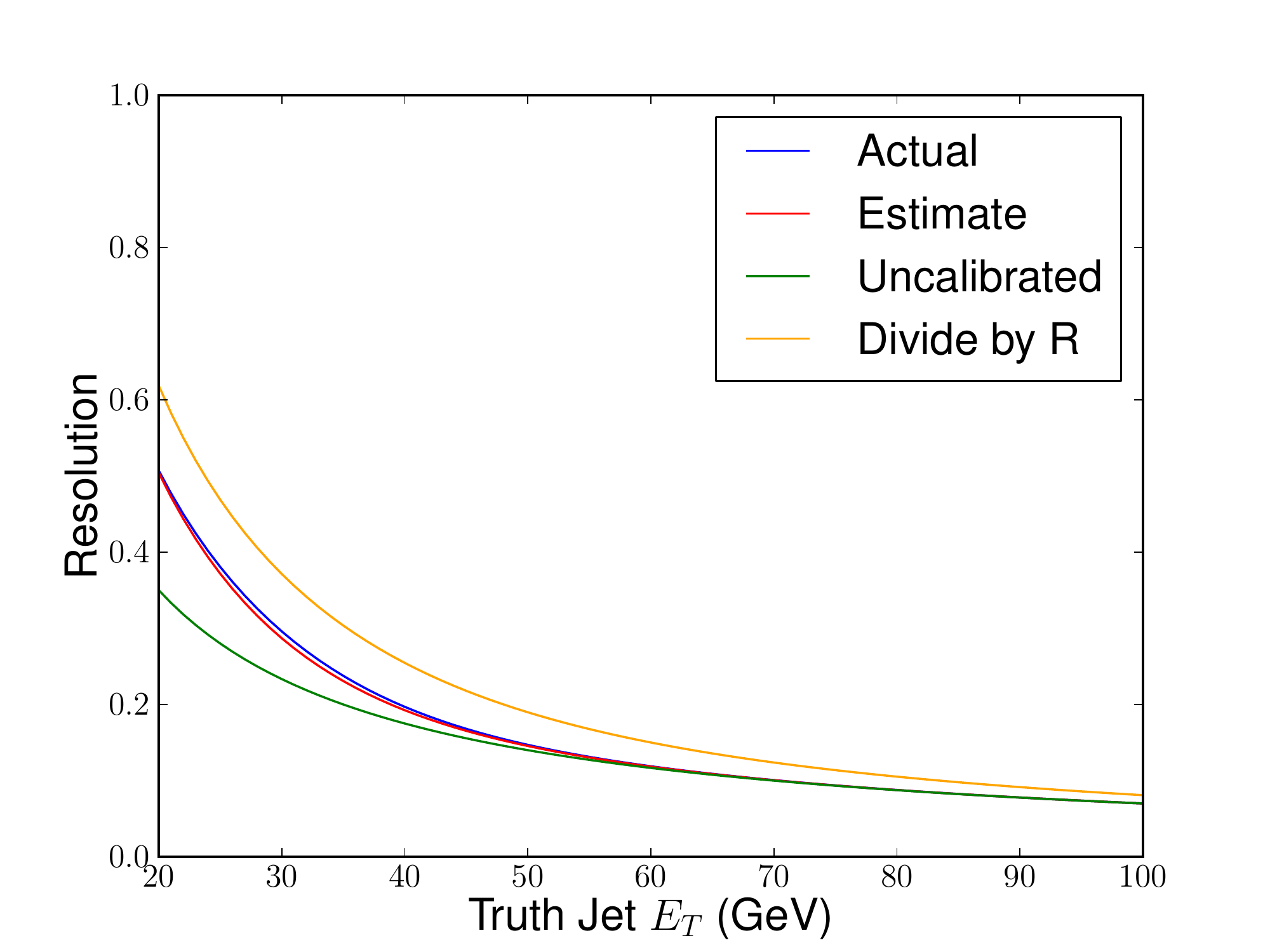}
\end{center}
\caption{The resolution of the $E_\text{T}$ distribution following numerical inversion when using the mean to calibrate, using a toy model similar to conditions in ATLAS or CMS. In blue, the exact calculated resolution. In red, the estimate of the closure using the first term of the higher-order correction in Eq.~\ref{eqn:resolutionseries_text}. In green, the uncalibrated resolution. In orange, the resolution when dividing by the response $R(x)$. For details of the model, see~\ref{sec:toy_model}.}
\label{fig:mean_resolution}
\end{figure}

The final calibrated resolution of the reconstructed jets is defined to be the standard deviation of the $Z$ distribution, with $X=x$, which is given by
\begin{align}
\hat{\sigma}(x)^2\equiv\sigma\left(Z|X=x\right)^2 \equiv \mathbb{E}\left[Z^2\middle| X=x\right]-\mathbb{E}\left[Z\middle| X=x\right]^2,
\end{align}
and the fractional resolution is just given by $\sigma\left(\frac{Z}{x}|X=x\right)$.  The fractional resolution, to first order in the Taylor series, is given by
\begin{align}
\sigma\left(\frac{Z}{x}|X=x\right)=\frac{1}{x}\hat{\sigma}(x) \approx \frac{1}{x}\frac{\sigma(x)}{f'(x)},
\label{eqn:resolutionseries_text}
\end{align}
the derivation of which can be found in~\ref{sec:calibrated_resolution_calculation}.  Note that $f'(x)$ is \emph{not} the response $R(x)=\frac{f(x)}{x}$. In particular, $f'(x)=R(x)+R'(x)x$, so $f'(x)\ne R(x)$ unless $R'(x)=0$, or equivalently $f(x)= kx$ for some constant $k$ (which is not the case at ATLAS nor at CMS). Figure~\ref{fig:mean_resolution} verifies Eq.~\ref{eqn:resolutionseries_text} and compares it to the method of dividing the width of the distribution by $R$, which is a standard diagnostic technique when a full calibration is not applied.

\subsection{Mode}
\label{sec:mode}
In the following section only, for brevity, we will let $f$ be $f_\text{mo}$ and $C$ be $C_\text{mo}$.  The distribution $\rho_{Y|X}(y|x)$ is usually unimodal and Gaussian fits to the ``core'' of this distribution are essentially picking out the mode of the distribution.  Therefore, the results of this section are a good approximation to what is often used in practice. We note that in the case that the underlying distribution is multimodal, it is not clear how to unambiguously define the mode of the distribution, and so the results of this section cannot be applied naively.

\subsubsection{Closure}
\label{sec:modeclosuresection}
Assuming that the probability distribution function is unimodal, the mode is the point at which the first derivative of the function is 0:
\begin{align}
f(x) = y^* \text{ s.t. } \rho'_Y(y^*) = 0.
\label{eqn:modedef}
\end{align}
Then we can write the closure condition (Eq.~\ref{eqn:closure}) as
\begin{align}
\text{mode}\left[\frac{Z}{x}\middle| X=x\right] = 1\nonumber\\
\rightarrow \text{mode}\left[Z\middle| X=x\right] = x\nonumber\\
\rightarrow \rho'_Z(x) = 0.
\label{eqn:modeclosuredef}
\end{align}

\noindent Using this definition, we can prove a result similar (but stronger) to the closure condition for the mean in the previous section:

\vspace{5mm}

\noindent  {\it Let the notion of central tendency be the mode. Numerical inversion closes if and only if $f$ is linear.}

\vspace{5mm}

\noindent  {\bf Proof.} We have from Eq.~\ref{eqn:newdist} that
\begin{align}
\rho_Z(z) = f'(z)\rho_Y(f(z)).
\end{align}
Therefore,
\begin{align}
\rho'_Z(z) = f''(z)\rho_Y(f(z))+f'(z)^2\rho'_Y(f(z)),
\end{align}
and
\begin{align}
\rho'_Z(x) &= f''(x)\rho_Y(f(x))+f'(x)^2\rho'_Y(f(x))\nonumber\\
&=f''(x)\rho_Y(y^*)+f'(x)^2\rho'_Y(y^*)\nonumber\\
&=f''(x)\rho_Y(y^*),
\end{align}
where $\rho_Y(y^*)>0$ since $y^*$ is the mode of the distribution $\rho_Y$. Then we see that if $f''(x)=0$, then $\rho'_Z(x)=0$ and closure is achieved.  In contrast, if $f''(x)\ne 0$, then $\rho'_Z(x)\ne 0$ and closure is not achieved. $\hspace{0.5 cm}\Box$

\vspace{5mm}

\noindent The closure when using the mode to calibrate, to first order in the Taylor series, is given by
\begin{align}
C \approx 1+\frac{f''(x)}{f'(x)^3}\frac{\tilde{\sigma}(x)^2}{x},
\label{eqn:closure_mode_text}
\end{align}
where $\tilde{\sigma}(x)$ is the width of a Gaussian fitted to just the area near the peak of the function $\rho_{Y|X}(y|x)$ (defined precisely in the next section). The derivation of Eq.~\ref{eqn:closure_mode_text} can be found in~\ref{sec:calibrated_mode_calculation}.

\begin{figure}[]
\begin{center}
\includegraphics[width=0.9\textwidth]{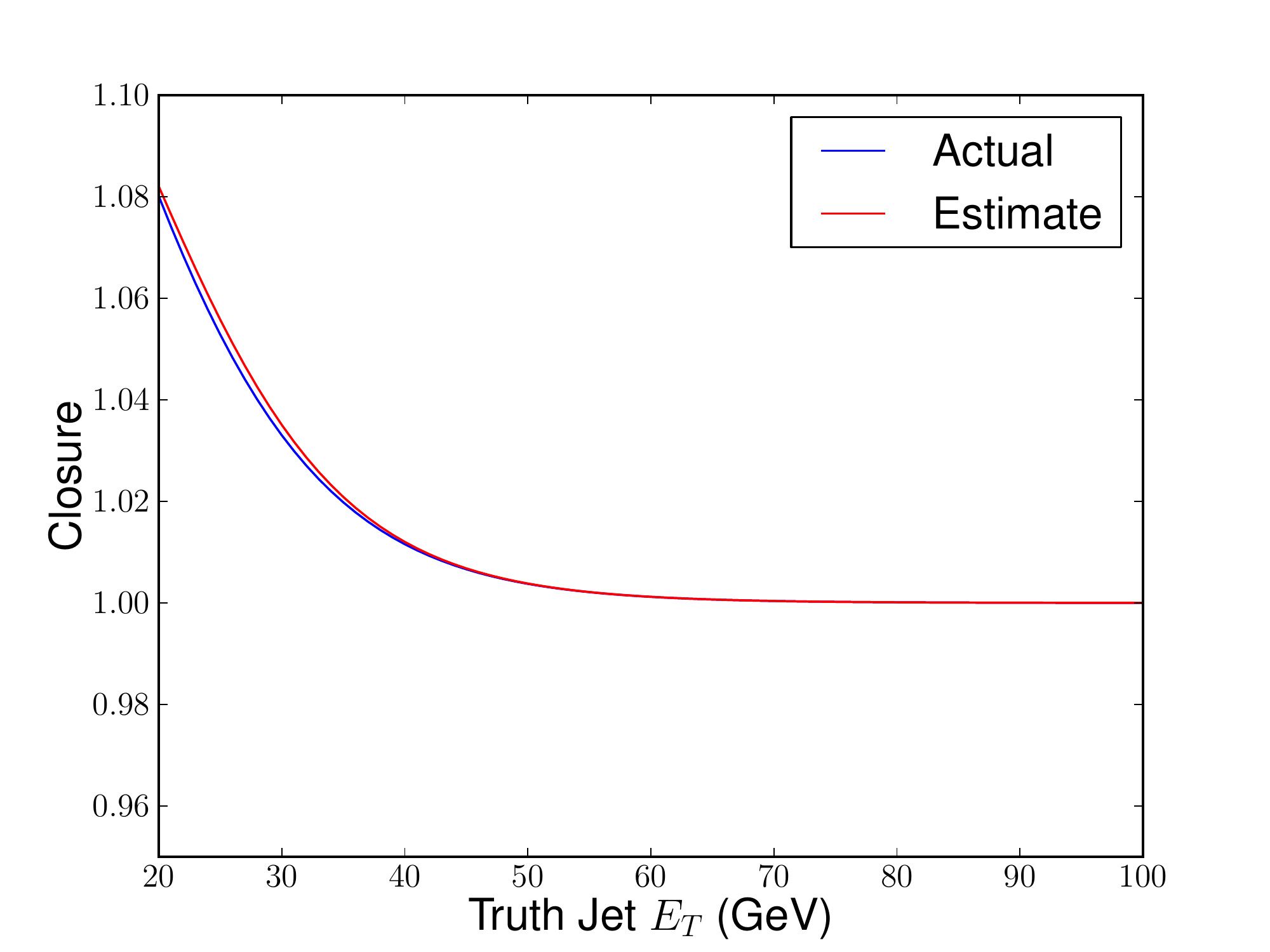}
\end{center}
\caption{The closure of numerical inversion when using the mode to calibrate, using a toy model similar to conditions in ATLAS or CMS. In blue, the exact calculated closure. In red, the estimate of the closure using the first term of the higher-order correction given in Eq.~\ref{eqn:closure_mode_text}. For details of the model, see~\ref{sec:toy_model}.}
\label{fig:mode_closure}
\end{figure}

Figure~\ref{fig:mode_closure} shows the inherent non-closure in numerical inversion for a toy calculation using a response function $R(x)$ that is typical for ATLAS or CMS, and the first term of the higher-order correction given in Eq.~\ref{eqn:closureseries_text}, when using the mode for calibration.

\subsubsection{Resolution}
Let $z^*(x)$ be the mode of the distribution $Z|X=x$, which is not necessarily equal to $x$ given the above result. It is often the case at ATLAS and CMS that a Gaussian is fit to the distributions $\rho_{Y|X}(y|x)$ and $\rho_{Z|X}(z|x)$ only in the vicinity of the modes $f(x)$ and $z^*(x)$, respectively, since it is assumed that the distributions have a Gaussian core but non-Gaussian tails. The width of the Gaussian core found in this fit is then used as a measure of the resolution of the distribution.  We thus define a ``trimmed resolution'' for a distribution $P$ with probability distribution function $\rho_P(p)$ about its mode $m$, which is valid if $P\sim\mathcal{N}(m,\tilde{\sigma})$ for $p$ near $m$:
\begin{align}
\label{eq:tildesigma}
\tilde{\sigma}(P)^2 \equiv -\frac{\rho_P(m)}{\rho_P''(m)}.
\end{align}

\noindent The definition in Eq.~\ref{eq:tildesigma} is chosen because it reduces to the usual variance for a Gaussian distribution.  For the distributions $\rho_{Y|X}(y|x)$ and $\rho_{Z|X}(z|x)$, we thus have the trimmed resolutions
\begin{align}
\tilde{\sigma}(x)^2\equiv\tilde{\sigma}\left(Y|X=x\right)^2 = -\frac{\rho_Y(f(x))}{\rho_Y''(f(x))} \\
\hat{\tilde{\sigma}}(x)^2\equiv\tilde{\sigma}\left(Z|X=x\right)^2 = -\frac{\rho_Z(z^*(x))}{\rho_Z''(z^*(x))}.
\label{mode_resolution_def}
\end{align}

\noindent The calibrated fractional trimmed resolution $\tilde{\sigma}\left(\frac{Z}{x}|X=x\right)$, to first order in the Taylor series, is given by
\begin{align}
\tilde{\sigma}\left(\frac{Z}{x}|X=x\right) = \frac{1}{x}\hat{\tilde{\sigma}}(x) \approx \frac{1}{x}\frac{\tilde{\sigma}(x)}{f'(x)},
\label{eqn:resolutionmode_text}
\end{align}
the derivation of which can be found in~\ref{sec:mode_resolution_calculation}.

\subsection{Median}
\label{sec:median}
In the previous sections we have examined using the mean or the mode to define $f$ and $C$, and found that both results do not lead to closure in general. We propose a new definition, using the median of the reconstructed jet $E_\text{T}$ distributions:
\begin{align}
f_\text{med}(x)&=\text{median}[Y|X=x]\\
R_\text{med}(x) &= \text{median}\left[\frac{Y}{x}\middle| X=x\right] = \frac{f_\text{med}(x)}{x}. 
\end{align}
And define $C_\text{med}$ analogously.  In the following section only, for brevity, we will let $f$ be $f_\text{med}$ and $C$ be $C_\text{med}$.
\subsubsection{Closure}
\label{sec:medianclosuresection}
The median of the distribution is the point at which 50\% of the distribution is above and 50\% is below:
\begin{align}
f(x) = y^* \text{ s.t. } \int_{-\infty}^{y^*} \rho_Y(y) dy = 0.5.
\end{align}
Then the closure condition (Eq.~\ref{eqn:closure}) can be written
\begin{align}
\text{median}\left[\frac{Z}{x}\middle| X=x\right] = 1\nonumber\\
\rightarrow \text{median}\left[Z\middle| X=x\right] = x\nonumber\\
\rightarrow \int_{-\infty}^{x} \rho_Z(z) dz = 0.5.
\label{eqn:medianclosuredef}
\end{align}
We can see then the following result under this definition of central tendency:

\vspace{5mm}

\noindent {\it Let the notion of central tendency be the median. Then numerical inversion always closes.}

\vspace{5mm}

\noindent {\bf Proof.} We have from Eq.~\ref{eqn:newdist} that
\begin{align}
\rho_Z(z) = f'(z)\rho_Y(f(z)).
\end{align}
So the closure condition in Eq.~\ref{eqn:medianclosuredef} becomes
\begin{align}
0.5 &= \int_{-\infty}^{x} \rho_Z(z) dz\nonumber\\
&=\int_{-\infty}^{x} f'(z)\rho_Y(f(z)) dz.
\end{align}
Then with $u = f(z), du = f'(z) dz$ we have
\begin{align}
0.5 &= \int_{-\infty}^{f(x)} \rho_Y(u) du\nonumber\\
&=\int_{-\infty}^{y^*} \rho_Y(u) du\nonumber\\
&=0.5. \hspace{1 cm} \Box
\end{align}

\subsubsection{Resolution}
A natural definition of resolution when using the median to calibrate jets is the 68\% interquantile range, defined as follows for a distribution $P$ with probability density function $\rho_P(p)$:

\noindent With $I_P^-$ and $I_P^+$ defined by
\begin{align}
\int_{-\infty}^{I_P^-}\rho_P(p)dp \equiv \Phi(-1),\\
\int_{-\infty}^{I_P^+}\rho_P(p)dp \equiv \Phi(+1);
\end{align}
the 68\% interquantile range is defined as
\begin{align}
\sigma_\text{IQR}(P) \equiv \frac{1}{2}\left(I_P^+-I_P^-\right).
\end{align}
Where $\Phi(x)=\frac{1}{2}\text{erfc}\left(\frac{-x}{\sqrt{2}}\right)$ is the cumulative distribution function of the normal distribution. The definition is designed so that if $P\sim\mathcal{N}(\mu,\sigma)$ then $\sigma_\text{IQR}(P)=\sigma$. The quantity $\sigma_\text{IQR}$ is called the ``68\% interquantile range'' because $\Phi(+1)-\Phi(-1) \approx 0.68$.  For the distributions $Y|X=x$ and $Z|X=x$, define:

\begin{align}
\sigma_\text{IQR}(x) = \sigma_\text{IQR}(Y|X=x)\\
\hat{\sigma}_\text{IQR}(x) = \sigma_\text{IQR}(Z|X=x).
\end{align}

\noindent Then we can see the following result for the calibrated resolution $\sigma_\text{IQR}(\frac{Z}{x}|X=x)$:

\vspace{5mm}

\noindent {\it The 68\% IQR of the calibrated response distribution is given by $\sigma_\text{IQR}(\frac{Z}{x}|X=x) = \frac{1}{2x}\left(f^{-1}(I_Y^+)-f^{-1}(I_Y^-)\right)$.}

\vspace{5mm}

\noindent {\bf Proof.}
We have
\begin{align}
\int_{-\infty}^{I_Z^-}\rho_Z(z)dz = \Phi(-1)\\
\int_{-\infty}^{I_Z^+}\rho_Z(z)dz = \Phi(+1).
\end{align}
From Eq.~\ref{eqn:newdist},
\begin{align}
\rho_Z(z) = f'(z)\rho_Y(f(z)),
\end{align}
so that
\begin{align}
\Phi(-1) &= \int_{-\infty}^{I_Z^-}f'(z)\rho_Y(f(z))dz\nonumber\\
&=\int_{-\infty}^{f(I_Z^-)}\rho_Y(u)du\nonumber\\
\rightarrow f(I_Z^-) &= I_Y^-\\
\Phi(+1) &= \int_{-\infty}^{I_Z^+}f'(z)\rho_Y(f(z))dz\nonumber\\
&=\int_{-\infty}^{f(I_Z^+)}\rho_Y(u)du\nonumber\\
\rightarrow f(I_Z^+) &= I_Y^+.
\end{align}
Therefore,
\begin{align}
\sigma_\text{IQR}(Z|X=x) &= \frac{1}{2}\left(I_Z^+-I_Z^-\right)\nonumber\\
&=\frac{1}{2}\left(f^{-1}(I_Y^+)-f^{-1}(I_Y^-)\right),
\end{align}
and
\begin{align}
\sigma_\text{IQR}\left(\frac{Z}{x}|X=x\right)=\frac{1}{2x}\left(f^{-1}(I_Y^+)-f^{-1}(I_Y^-)\right).\hspace{1 cm}\Box
\label{eqn:resolutionmedian_text}
\end{align}
\newpage
\section{Discussion}
\label{sec:discussion}

After a quick summary in Section~\ref{sec:summary} of the results presented so far, Section~\ref{sec:recommendations} discusses the benefits and drawbacks of various methods of calibration, and Sections~\ref{sec:iterated_text} and~\ref{sec:corrected_numerical_inversion_text} describe extensions of numerical inversion that may help to improve closure.

\subsection{Summary of Results}
\label{sec:summary}
In Section~\ref{sec:introclosure} we defined the concept of closure in the process of calibrating the $E_\text{T}$ of jets.  We found in Sections~\ref{sec:meanclosuresection} and ~\ref{sec:modeclosuresection} that when using the mean or mode, respectively, of the distribution $Y|X=x$ to calibrate, closure is not necessarily achieved; with the response functions found at ATLAS or CMS, it is expected that numerical inversion will not close. We also provided estimates for the non-closure for the mean (Eq.~\ref{eqn:closureseries_text}) and for the mode (Eq.~\ref{eqn:closure_mode_text}). In those estimates we find that as the underlying resolution $\sigma(x)$ or $\tilde{\sigma}(x)$ of the uncalibrated jet distribution $Y|X=x$ increases, the non-closure gets worse. This indicates that the non-closure issues raised in this note will become more important as the LHC moves to conditions with higher pileup in the future.

\vspace{2mm}

A new calibration scheme based on the median of $Y|X=x$ is proposed in Section~\ref{sec:medianclosuresection}.  With this method of calibration, closure is always achieved.

\vspace{2mm}

Each section also explored various definitions of the resolution of the fractional calibrated jet distribution $\frac{Z}{x}|X=x$, where the most natural definition depends on the manner in which calibration has been performed (i.e., whether using the mean, mode, or median to calibrate). We provided useful estimates for the standard deviation (Eq.~\ref{eqn:resolutionseries_text}), the trimmed Gaussian width (Eq.~\ref{eqn:resolutionmode_text}), and an exact formula for the 68\% IQR (Eq.~\ref{eqn:resolutionmedian_text}). These expressions can be used to quickly estimate the final resolution of a jet algorithm without having to actually apply the calibration jet-by-jet.

\subsection{Recommendation for Method of Calibration}
\label{sec:recommendations}
As mentioned in the summary above, we have that for a non-linear response function closure is not necessarily achieved when using the mode or mean to calibrate, and closure is necessarily achieved when using the median. While this indicates that the median is a useful metric to use if closure is the main objective, we accept that there might be reasons to use the mode instead (for example, if the tails of $\rho_{Y|X}(y|X=x)$ are cut off, then the mode should stay constant while the median and mean will change). Thus we leave it to the reader to decide which method of calibration is most appropriate to use for their specific purposes. To that end, we also have discussion below about methods to improve the closure when the mode is being used to calibrate.

\subsection{Iterated Numerical Inversion}
\label{sec:iterated_text}
A natural question is whether it is useful for the purposes of achieving closure to implement numerical inversion again on the calibrated jet collection, if closure has not been achieved the first time. We define the \emph{iterated numerical inversion} process as follows:

\vspace{5mm}

\noindent With $C(x)$ defined as in Eq.~\ref{eqn:closure}, let
\begin{align}
R_{\text{new}}(x) &\equiv C(x)\\
f_{\text{new}}(x) &\equiv C(x)x.
\end{align}
Then, apply numerical inversion on the calibrated distribution $Z$:
\begin{align}
Z\mapsto Z_\text{new} = f_{\text{new}}^{-1}(Z).
\end{align}
We then ask if the closure of this new distribution, $C_\text{new}(x)$ (defined analogously as in Eq.~\ref{eqn:closure}), is closer to 1 than $C(x)$.  In general, this is a difficult question to answer, but we have derived analytic approximations when the mode is used to derive the calibration (see~\ref{sec:iterated}).  Iterating numerical inversion does \emph{not} always help:

\begin{align}
\frac{|C_\text{new}(x)-1|}{|C(x)-1|} &\approx \frac{12f''(x)^2\tilde{\sigma}(x)^2}{f'(x)^4}\label{eqn:iterated_closure_ratio}.
\end{align}
If the ratio in Eq.~\ref{eqn:iterated_closure_ratio} is greater than 1, then the closure gets worse after a second iteration of numerical inversion.  In particular, as $\tilde{\sigma}(x)$ gets larger, the iterated closure gets worse relative to the original closure. So we expect at higher levels of pileup that iterating numerical inversion will not be useful. In Figure~\ref{fig:mode_closure_bigs} we can see that iterating numerical inversion does make the closure worse than the original closure, in a model simulating higher pileup conditions.  The next section provides another scheme to correct for the residual non-closure that does not require iterating the process of numerical inversion.

\subsection{Corrected Numerical Inversion}
\label{sec:corrected_numerical_inversion_text}
As noted above, when using the mean or mode of the distribution $Y|X=x$ to calibrate, closure is not achieved in general. With the closed-form estimates of the non-closure provided in the text, one might think to simply ``subtract off'' the non-closure. However the non-closure estimates provided are in terms of the truth $E_\text{T}$ value $x$.  Since $x$ is not available in data, a sensible proxy is to use numerical inversion as an estimate for $x$.  This is actually equivalent to iterated numerical inversion, which as shown in the previous section does not always help.

Another possibility is to use a different original response function to perform the calibration.  Suppose that instead of using $f(x)=R(x)x$, there was a new function $g(x)\ne f(x)$ such that if the calibration is performed with this new function, $Y\mapsto Z_\text{corr}=g^{-1}(Y)$, the new calibrated distribution $Z_\text{corr}|X=x$ does achieve closure or gets closer to achieving closure than when calibrating using $f$.

We define the \emph{corrected numerical inversion} process as follows:
\begin{enumerate}
\item Calculate $f(x)=f_\text{mo}(x)=\text{mode}[Y|X=x]$.
\item Let $g(x) = g(x;f(x))$ be a calibration function depending on the fitted function $f(x)$.
\item Apply the calibration $Y\mapsto Z_\text{corr}=g^{-1}(Y)$ jet-by-jet.
\end{enumerate}
We then can examine the closure
\begin{align}
C_\text{corr}(x) = \text{mode}\left[\frac{Z_\text{corr}}{x}\Big|X=x\right].
\end{align}
And say we have achieved closure if
\begin{align}
C_\text{corr}(x) \equiv 1.
\end{align}
We examine the case of using the mode to measure closure, again because in practice that is what is often used when there are significant non-Gaussian tails. 

One way to specify $g$ is by explicitly requiring closure. In~\ref{sec:corrected_numerical_inversion_calculation} it is shown that in the case that closure is achieved exactly, $g$ necessarily satisfies the differential equation\footnote{As noted in the derivation, this equation also assumes the following: that the underlying distribution $Y|X=x$ is approximately Gaussian in the vicinity of its mode $f(x)$; and that the correction is small, with $|g(x)-f(x)|\ll \tilde{\sigma}(x)$.}
\begin{align}
0=g''(x)-g'(x)^2\frac{g(x)-f(x)}{\tilde{\sigma}(x)^2}.
\label{eqn:diffeq}
\end{align}
In principle Eq.~\ref{eqn:diffeq} can be solved numerically given numerical fitted values $f(x)$ and $\tilde{\sigma}(x)$, though in practice such a method may prove intractable.

\begin{figure}[]
\begin{center}
\includegraphics[width=0.9\textwidth]{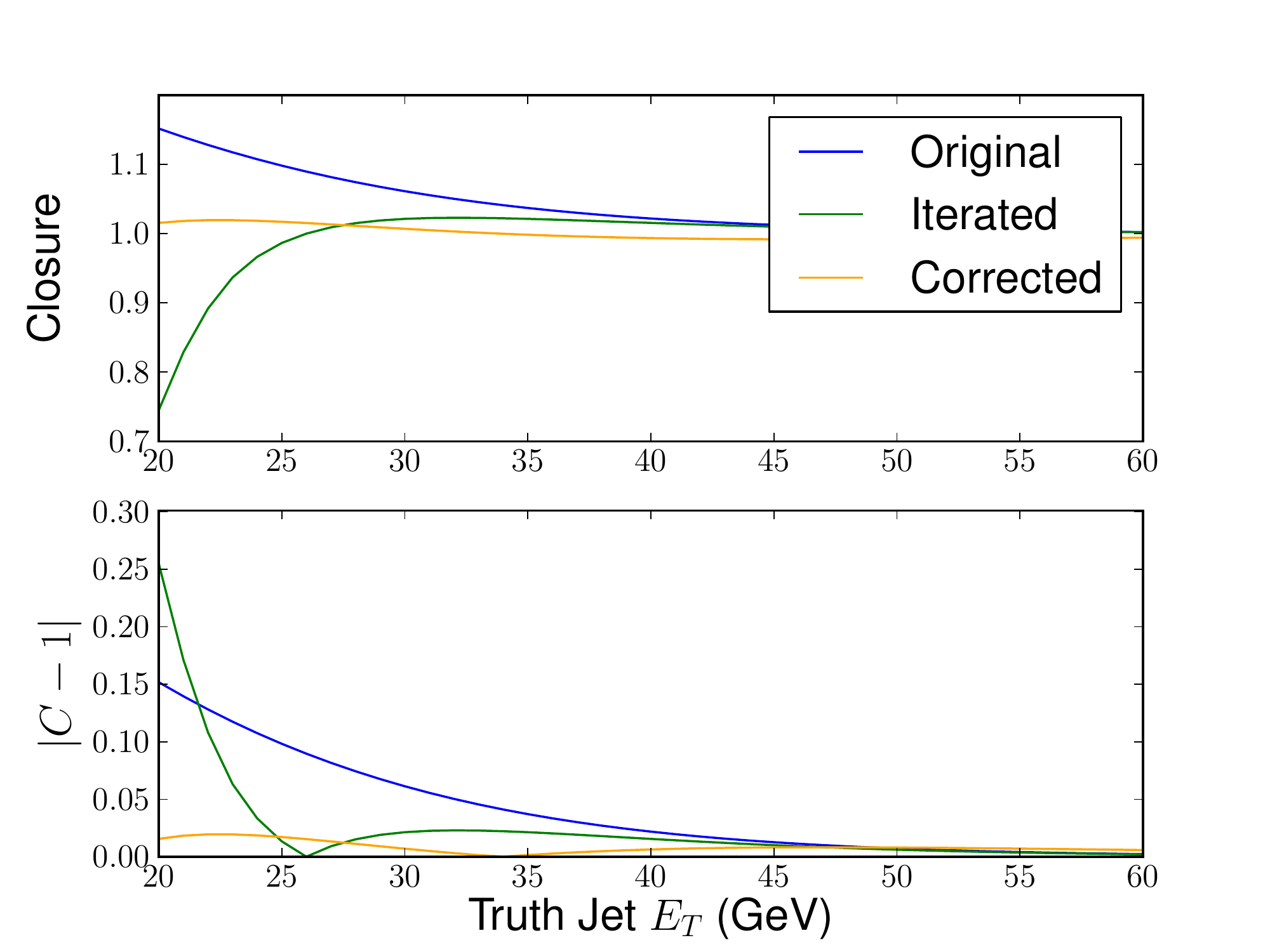}
\end{center}
\caption{The top plot shows the closure of numerical inversion when using the mode to calibrate, using a toy model similar to conditions in ATLAS or CMS but increasing $\sigma(x)$ by a factor of 1.4 in order to simulate higher pileup conditions. In blue, the original closure as defined in Eq.~\ref{eqn:closure}. In green, the closure after iterating numerical inversion once as in Section~\ref{sec:iterated_text}. In orange, the closure after using the parameterized corrected numerical inversion technique as in Section~\ref{sec:corrected_numerical_inversion_text}. For details of the model, see~\ref{sec:toy_model}. The bottom plot shows the absolute non-closure $|C-1|$. In particular, at low $E_\text{T}$, iterating numerical inversion does worse, while corrected numerical inversion does better than the original calibration.}
\label{fig:mode_closure_bigs}
\end{figure}

Another way to specify $g$ is to use external parameters
\begin{align}
g(x) = g(x;f(x);a_1,...,a_n).
\end{align}
Then the parameters $a_1,...,a_n$ can be chosen such that the closure is as close to 1 as possible. This is the method used to find the corrected calibration curve in Figure~\ref{fig:mode_closure_bigs}, and explained in more detail in~\ref{sec:corrected_numerical_inversion_parameterization}. The absolute non-closure $|C-1|$ is significantly smaller than the original non-closure, even in a model simulating very high pileup conditions.

\section{Conclusions}
\label{sec:conclusion}

Jets are ubiquitous at the LHC and their calibration is one of the most important preprocessing steps for data analysis.  The standard technique for jet calibration is numerical inversion.  This paper has formally defined numerical inversion and derived many of its properties.  The three most important results:

\begin{itemize}
\item {\bf Numerical inversion is inherently biased}: calibrated reconstructed jets are not guaranteed to be centered around the corresponding particle-level jet $E_\text{T}$.  However, when the median is used for the notion of `centered', closure is guaranteed.  In practice where the detector response is non-linear, there is never closure when `centered' means the mode of the response distribution.
\item {\bf Numerical inversion can be approximated}: However, the resolution of the calibrated jets is not well-approximated by the uncalibrated jet resolution divided by the response.  Calibrated resolutions can still be simply estimated, but they depend on the derivative of the calibration function and not the response.
\item {\bf Numerical inversion can be improved}: Modified calibration functions can be constructed to achieve a better closure than using the same measure of central tendency for deriving the calibration function and assessing the closure.
\end{itemize}

These considerations may become even more important in the future when fluctuations in the detector response increase due to the presence of larger contributions from pileup.  Numerical inversion is a general technique that can be applied to any detector calibration where a reliable simulation exists for matching objects before and after the detector response.  The results presented here may therefore have a broader applicability than to jets, the LHC, or even high energy physics.

\section{Acknowledgments}

We would like to thank Alexx Perloff for his detailed comments on the manuscript and for useful discussions about the CMS jet calibration procedure.  Additionally, we acknowledge Francesco Rubbo for his helpful suggestions on the manuscript.  For a portion of this work, BN was supported by the NSF Graduate Research Fellowship under Grant No. DGE-4747 and by the Stanford Graduate Fellowship.  

\clearpage
\newpage

\appendix

\section{Gaussian Invariance Lemma}
\label{sec:lemma}

{\it Let $X\sim \mathcal{N}(\mu,\sigma)$ and $f$ be some function such that $f'(x)>0$.  Then, $f(X)\sim\mathcal{N}(\mu',\sigma')$ if and only if $f(x)$ is linear in $x$.}

\vspace{5mm}

\noindent {\bf Proof.} The converse is a well-known result, and can be obtained directly from application of Eq.~\ref{eqn:newdist}.

Now suppose that $f(X)\sim\mathcal{N}(\mu',\sigma')$.  Let $Y=(X-\mu)/\sigma$ and define 
\begin{align}
g(y)=\frac{f(\sigma y+\mu)-\mu'}{\sigma'},
\end{align}
so that $Y$ and $Z=g(Y)$ both have a standard normal distribution. Furthermore,
\begin{align}
g'(y) = \frac{\sigma}{\sigma'}f'(\sigma y+\mu) > 0,
\end{align}
so $g$ is monotonic.

We then can write for any $c$:
\begin{align}
\nonumber
\Phi(c)=\Pr(Y<c)&=\Pr(g(Y)<g(c))\\\nonumber
&=\Pr(Z<g(c))\\
&=\Phi(g(c)),
\end{align}
Where $\Phi(x)$ is the normal distribution cumulative distribution function. Since $\Phi$ is invertible, we then have that $g(c)=c$.  Inserting the definition of $g$ then gives us the final result:
\begin{align}
f(x)=\frac{\sigma'}{\sigma} (x-\mu)+\mu'.\hspace{1 cm}\Box
\end{align}

\section{Closure of the Mean}
\label{sec:mean_nonclosure}
{\it The closure of jets reconstructed from truth jets with $E_\text{T} = x$ and $f(x)=f_{me}(x)$ is given to first order by $C\approx 1-\frac{1}{2}\frac{f''(x)}{f'(x)^3}\frac{\sigma(x)^2}{x}$.}

\vspace{5mm}

\noindent  {\bf Derivation.}
We begin by Taylor expanding $f^{-1}(y)$ about $y=f(x)$:
\begin{align}
f^{-1}(y) &= \sum_{n=0}^\infty \frac{1}{n!}\left(f^{-1}\right)^{(n)}\left(f(x)\right)\cdot\left(y-f(x)\right)^n\nonumber\\
&=\sum_{n=0}^\infty \frac{1}{n!}g_n(x)\cdot\left(y-f(x)\right)^n,
\end{align}
where $g_n(x) \equiv (f^{-1})^{(n)}(f(x))$ means the $n$th derivative of $f^{-1}(y)$, evaluated at $y=f(x)$. Plugging this into Eq.~\ref{eqn:closuredef}, we have
\begin{align}
C &= \frac{1}{x}\int dy \rho_{Y|X}(y|x) f^{-1}(y)\nonumber\\
&=\sum_{n=0}^\infty \frac{1}{n!}\frac{g_n(x)}{x}\int dy \rho_{Y|X}(y|x) \left(y-f(x)\right)^n\nonumber\\
&=\sum_{n=0}^\infty \frac{1}{n!}\frac{g_n(x)}{x} \mu_n(x),
\end{align}
where $\mu_n(x)$ are the standard central moments $\mu_n(x) = \mathbb{E}\left[\left(Y-\mathbb{E}\left[Y\right]\right)^n\middle| X=x\right]$, since by definition $f(x)=\mathbb{E}[Y|X=x]$.

The first few central moments are independent of the distribution $\rho_{Y|X}$.  In particular, $\mu_0 = 1$ is the normalization, and $\mu_1 = 0$. Writing these terms out, we have
\begin{align}
C =\frac{g_0(x)}{x}+\sum_{n=2}^\infty \frac{1}{n!}\frac{g_n(x)}{x} \mu_n(x).
\end{align}
Noting that $g_0(x) = f^{-1}(f(x)) = x$,
\begin{align}
C &=1+\sum_{n=2}^\infty \frac{1}{n!}\frac{g_n(x)}{x} \mu_n(x).\label{eqn:closureseries}
\end{align}

\noindent We see that, if $f$ is linear, then so is $f^{-1}$, and so $g_n = 0$ for all $n\ge 2$. Then Eq.~\ref{eqn:closureseries} reduces to $C=1$, and numerical inversion closes, as was found in Eq.~\ref{eqn:closure_linear_proof}.

It will be instructive to expand out the first few terms of Eq.~\ref{eqn:closureseries}. We note that, by definition, $\mu_2(x) = \sigma(x)^2$ is the variance, and $\mu_3(x) = \sigma(x)^3\gamma_1$ defines the skew $\gamma_1$. Then we have
\begin{align}
C &=1+\frac{1}{2}\frac{g_2(x)}{x}\sigma(x)^2+\frac{1}{6}\frac{g_3(x)}{x}\sigma(x)^3\gamma_1+\sum_{n=4}^\infty \frac{1}{n!}\frac{g_n(x)}{x} \mu_n(x).\label{eqn:closureseriesexpand}
\end{align}

Suppose we are given an arbitrary distribution specified by its moments $\mu_n(x)$. Then the requirement that closure is satisfied in the form of the right hand side of Eq.~\ref{eqn:closureseries} converging to $1$ exactly imposes strict constraints on the function $g(x)$, so that only for a highly specific choice of $g$ and therefore $f$ is closure achieved. Thus in general we do not expect closure to be satisfied for an arbitrary initial distribution $\rho_{Y|X}$.

We note that, since we expect the derivatives $g_n(x)$ and the moments $\mu_n(x)$ to grow considerably slower than $n!$ for functions $f$ and distributions $\rho_{Y|X}$ encountered at the LHC, we expect Eq.~\ref{eqn:closureseries} to converge, and Eq.~\ref{eqn:closureseriesexpand} gives the dominant contributions to the non-closure, i.e.
\begin{align}
C \approx 1+\frac{1}{2}\frac{g_2(x)}{x}\sigma(x)^2+\frac{1}{6}\frac{g_3(x)}{x}\sigma(x)^3\gamma_1.
\end{align}

If $\rho_{Y|X}$ is symmetric or near-symmetric, or if the third derivative of $g$ is small, such that $g_3(x)\sigma(x)\gamma_1 \ll g_2(x)$, then the dominant contribution to the non-closure is just
\begin{align}
C \approx 1+\frac{1}{2}\frac{g_2(x)}{x}\sigma(x)^2.
\end{align}

\noindent We further note that
\begin{align}
g_2(x) &= (f^{-1})^{(2)}(f(x)) = -\frac{f''(x)}{f'(x)^3}\nonumber\\
\rightarrow C &\approx 1-\frac{1}{2}\frac{f''(x)}{f'(x)^3}\frac{\sigma(x)^2}{x}.\hspace{1 cm}\Box
\label{eqn:closureseriesgaussian}
\end{align}

\newpage
\section{Calibrated Resolution of the Mean}
\label{sec:calibrated_resolution_calculation}
{\it The calibrated resolution of jets reconstructed from truth jets with $E_\text{T} = x$ and $f(x)=f_{me}(x)$ is given to first order by $\frac{\sigma(x)}{f'(x)}$.}

\vspace{5mm}

\noindent {\bf Derivation.}
We note that, expanding $f^{-1}(y)$ about $y=f(x)$ out to one derivative, and using the definitions of $g_n(x)$ and $\mu_n(x)$ from the previous section,
\begin{align}
(f^{-1}(y))^2 \approx g_0(x)^2+2g_0(x)g_1(x)(y-f(x))+g_1(x)^2(y-f(x))^2,
\end{align}
so that
\begin{align}
\mathbb{E}\left[Z^2\middle| X=x\right]&=\int dy \rho_{Y|X}(y|x) (f^{-1}(y))^2\nonumber\\
&\approx\int dy \rho_{Y|X}(y|x) \left(g_0(x)^2+2g_0(x)g_1(x)(y-f(x))+g_1(x)^2(y-f(x))^2\right)\nonumber\\
&=g_0(x)^2\mu_0(x)+2g_0(x)g_1(x)\mu_1(x)+g_1(x)^2\mu_2(x)\nonumber\\
&=g_0(x)^2+g_1(x)^2\sigma(x)^2.\hspace{5mm}\text{($\mu_1=0$ by construction)}
\end{align}
Out to one derivative we also have that (as derived in the previous section)
\begin{align}
\mathbb{E}\left[Z\middle| X=x\right]^2 &\approx g_0(x)^2\nonumber\\
\rightarrow \sigma\left(Z|X=x\right)^2 &= \mathbb{E}\left[Z^2\middle| X=x\right]-\mathbb{E}\left[Z\middle| X=x\right]^2\nonumber\\
&\approx g_1(x)^2\sigma(x)^2.
\end{align}
Then,
\begin{align}
g_1(x) = (f^{-1})'(f(x)) &= \frac{1}{f'(x)}\nonumber\\
\rightarrow \sigma\left(Z|X=x\right)^2 &\approx \frac{\sigma(x)^2}{f'(x)^2}\nonumber\\
\rightarrow \hat{\sigma}(x)=\sigma\left(Z|X=x\right) &\approx \frac{\sigma(x)}{f'(x)}. \hspace{1 cm} \Box \label{eqn:resolution}
\end{align}

\newpage
\section{Closure of the Mode}
\label{sec:calibrated_mode_calculation}
{\it The closure of jets reconstructed from truth jets with $E_\text{T} = x$ and $f(x)=f_{mo}(x)$ is given to first order by $C\approx 1+\frac{f''(x)}{f'(x)^3}\frac{\tilde{\sigma}(x)^2}{x}$.}

\vspace{5mm}

\noindent {\bf Derivation.}
As a reminder for the reader, for brevity, we will let $\rho_Y(y)=\rho_Y(y|x)$ and $\rho_Z(z)=\rho_Z(z|x)$, and let the parameter $x$ be understood.

We begin by supposing that the closure is not much different than 1, so that we can examine $\rho_Z(z)$ in the vicinity of $z=x$ to find the mode $z^*$. Expanding Eq.~\ref{eqn:newdist} about to second order in $(z-x)$:
\begin{align}
\rho_Z(z) &= f'(z)\rho_Y(f(z))\nonumber\\
&\approx \left[f'(x)+(z-x)f''(x)+\frac{(z-x)^2}{2}f'''(x)\right]\nonumber\\
&\times\left[\rho_Y(f(x))+(z-x)\rho_Y'(f(x))f'(x)+\frac{(z-x)^2}{2}\rho_Y''(f(x))f'(x)^2\right].
\end{align}
We note from the condition Eq.~\ref{eqn:modedef} that $\rho_Y'(f(x))=0$, so
\begin{align}
\rho_Z(z)&\approx \left[f'(x)+(z-x)f''(x)+\frac{(z-x)^2}{2}f'''(x)\right]\nonumber\\
&\times\left[\rho_Y(f(x))+\frac{(z-x)^2}{2}\rho_Y''(f(x))f'(x)^2\right]\nonumber\\
&\approx f'(x)\rho_Y(f(x))+(z-x)f''(x)\rho_Y(f(x))\nonumber\\
&+\frac{(z-x)^2}{2}\left[f'''(x)\rho_Y(f(x))+f'(x)^3\rho_Y''(f(x))\right],
\end{align}
so that
\begin{align}
\rho'_Z(z)&\approx f''(x)\rho_Y(f(x))+(z-x)\left[f'''(x)\rho_Y(f(x))+f'(x)^3\rho_Y''(f(x))\right].
\label{eqn:drhoz}
\end{align}
Then the closure condition Eq.~\ref{eqn:modeclosuredef} gives
\begin{align}
\rho'_Z(z^*)&=0\nonumber\\
\rightarrow z^* &\approx x-\frac{f''(x)\rho_Y(f(x))}{f'''(x)\rho_Y(f(x))+f'(x)^3\rho_Y''(f(x))},
\end{align}
i.e. the mode of $\rho_Z(z)$ occurs at $z=z^*$.  Then the closure is
\begin{align}
C &= \frac{z^*}{x}\nonumber\\
&\approx 1-\frac{1}{x}\frac{f''(x)\rho_Y(f(x))}{f'''(x)\rho_Y(f(x))+f'(x)^3\rho_Y''(f(x))}\nonumber\\
&=1-\frac{1}{x}\frac{f''(x)\frac{\rho_Y(f(x))}{\rho_Y''(f(x))}}{f'''(x)\frac{\rho_Y(f(x))}{\rho_Y''(f(x))}+f'(x)^3}\nonumber\\
&=1+\frac{f''(x)}{f'(x)^3-\tilde{\sigma}(x)^2f'''(x)}\frac{\tilde{\sigma}(x)^2}{x}.
\label{eqn:mode_closure_df3}
\end{align}

\begin{figure}[]
\begin{center}
\includegraphics[width=0.9\textwidth]{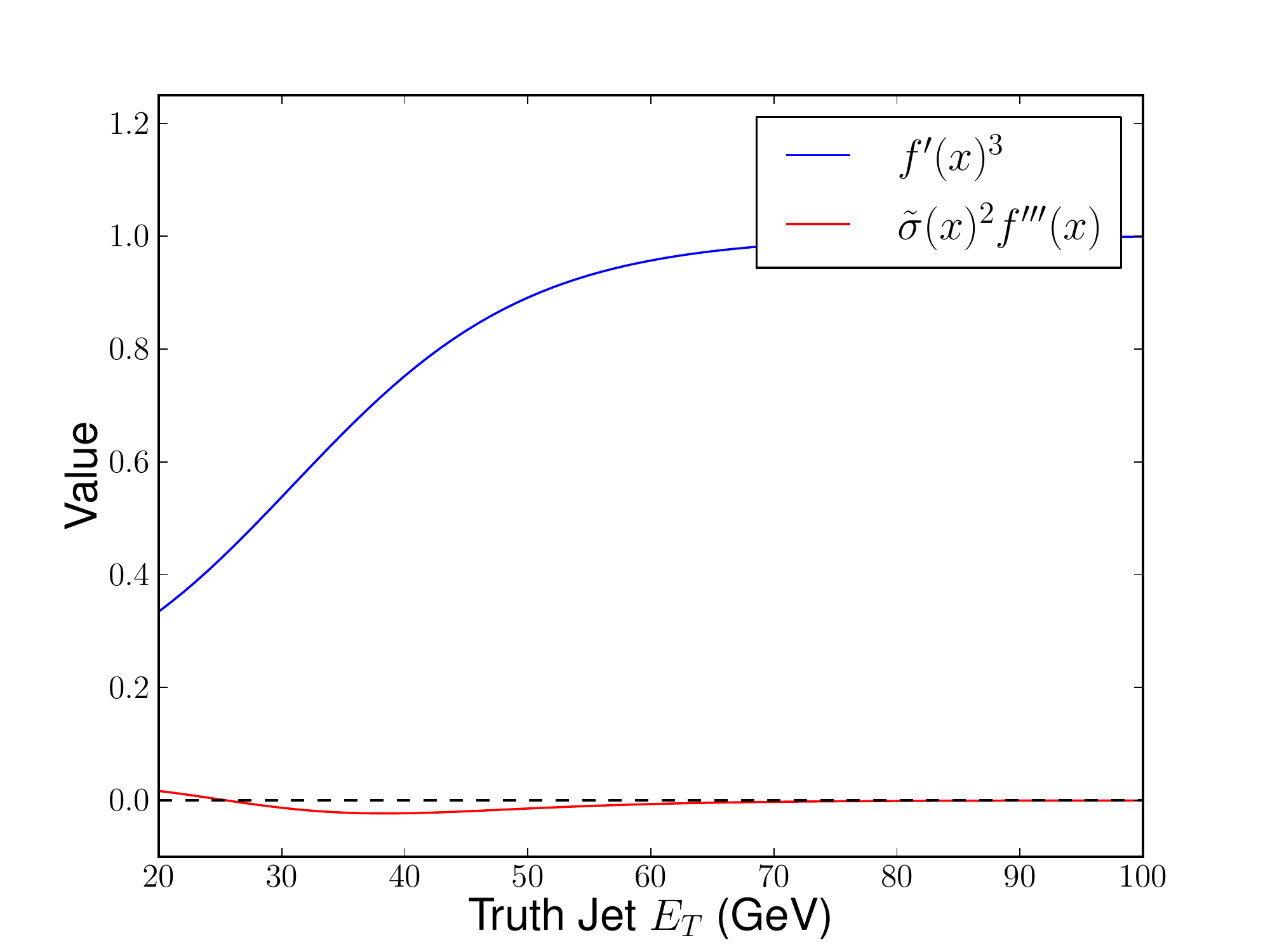}
\end{center}
\caption{A comparison of derivative values using a toy model similar to conditions in ATLAS or CMS. In blue, $f'(x)^3$. In red, $\tilde{\sigma}(x)^2f'''(x)$. For details of the model, see~\ref{sec:toy_model}.}
\label{fig:d_comp}
\end{figure}

In practice we find that for typical response functions, higher derivatives of $f$ tend to vanish. A comparison between the two terms in the denominator of Eq.~\ref{eqn:mode_closure_df3} can be found in Figure~\ref{fig:d_comp} for the toy model considered in~\ref{sec:toy_model}; we find that $f'(x)^3 \gg \tilde{\sigma}(x)^2f'''(x)$. Thus, in practice we recommend the approximation
\begin{align}
C\approx 1+\frac{f''(x)}{f'(x)^3}\frac{\tilde{\sigma}(x)^2}{x}.\hspace{1 cm} \Box
\label{eqn:mode_closure_simple}
\end{align}
The agreement between the actual and estimated closure in Figure~\ref{fig:mode_closure} also confirms this approximation. Thus, in the body of this text Eq.~\ref{eqn:mode_closure_simple} is presented as the result, even though Eq.~\ref{eqn:mode_closure_df3} is technically more precise.

\newpage
\section{Resolution of the Mode}
\label{sec:mode_resolution_calculation}
{\it The resolution of jets reconstructed from truth jets with $E_\text{T} = x$ and $f(x)=f_{mo}(x)$ is given to first order by $\hat{\tilde{\sigma}}(x)\approx \frac{\tilde{\sigma}(x)}{f'(x)}.$}

\vspace{5mm}

\noindent {\bf Derivation.}
From Eq.~\ref{eqn:drhoz} we have
\begin{align}
\rho''_Z(z)&\approx f'''(x)\rho_Y(f(x))+f'(x)^3\rho_Y''(f(x)).
\end{align}
Then the resolution is given as
\begin{align}
\hat{\tilde{\sigma}}(x)^2 &= -\frac{\rho_Z(z^*)}{\rho_Z''(z^*)}\nonumber\\
&\approx-\frac{f'(x)\rho_Y(f(x))}{f'''(x)\rho_Y(f(x))+f'(x)^3\rho_Y''(f(x))}\nonumber\\
&=\frac{f'(x)\tilde{\sigma}(x)^2}{f'(x)^3-f'''(x)\tilde{\sigma}(x)^2}.
\end{align}
Following the discussion in~\ref{sec:calibrated_mode_calculation}, we simplify the denominator to get the approximation
\begin{align}
\tilde{\sigma}(x)^2 &\approx \frac{\tilde{\sigma}(x)^2}{f'(x)^2}\nonumber\\
\rightarrow \tilde{\sigma}(x) &\approx \frac{\tilde{\sigma}(x)}{f'(x)}.\hspace{1 cm} \Box
\end{align}

\section{Iterated Numerical Inversion Calculation}
\label{sec:iterated}
{\it The closure $C_\text{new}(x)$ after iterating numerical inversion is not necessarily closer to 1 than the closure $C(x)$ after performing numerical inversion once.}

\vspace{5mm}

\noindent {\bf Derivation.}
We limit ourselves to the case that we are using the modes of the distributions $Y|X=x$ and $Z|X=x$ to calibrate, as in practice that is what is used at ATLAS and CMS for numerical inversion.

We use the estimation of the closure of the mode Eq.~\ref{eqn:closure_mode_text}:
\begin{align}
C(x) &\approx 1+\frac{f''(x)}{f'(x)^3}\frac{\tilde{\sigma}(x)^2}{x}\nonumber\\
\rightarrow |C(x)-1|&\approx \left|\frac{f''(x)}{f'(x)^3}\frac{\tilde{\sigma}(x)^2}{x}\right|.
\end{align}
We use the iterated numerical inversion response
\begin{align}
f_{\text{new}}(x) &= C(x)x\nonumber\\
&\approx x+\frac{f''(x)}{f'(x)^3}\tilde{\sigma}(x)^2\\
\rightarrow f'_{\text{new}}(x) &\approx 1-3\frac{f''(x)^2}{f'(x)^4}\tilde{\sigma}(x)^2\\
\rightarrow f''_{\text{new}}(x) &\approx 12\frac{f''(x)^3}{f'(x)^5}\tilde{\sigma}(x)^2.
\end{align}
Where we have ignored higher derivatives of $f(x)$\footnote{See, e.g., Figure~\ref{fig:d_comp}.} and derivatives of $\sigma(x)$\footnote{For this specific counterexample, we are examining the case that $\sigma'(x)=0$, which is realistic for high pileup conditions.}.  We also have the estimation of the resolution of the calibrated distribution Eq.~\ref{eqn:resolutionmode_text}
\begin{align}
\hat{\tilde{\sigma}}(x) \approx \frac{\tilde{\sigma}(x)}{f'(x)},
\end{align}

\noindent So that we can estimate the closure after iterating numerical inversion as
\begin{align}
C_\text{new}(x) &\approx 1+\frac{f''_\text{new}(x)}{f'_\text{new}(x)^3}\frac{\hat{\tilde{\sigma}}(x)^2}{x}\nonumber\\
&\approx 1+12\frac{f''(x)^3}{f'(x)^5}\tilde{\sigma}(x)^2\frac{\tilde{\sigma}(x)^2}{f'(x)^2}\frac{1}{x}\nonumber\\
&=1+\frac{12}{x}\frac{f''(x)^3}{f'(x)^7}\tilde{\sigma}(x)^4\\
\rightarrow |C_\text{new}(x)-1| &\approx \left|\frac{12}{x}\frac{f''(x)^3}{f'(x)^7}\tilde{\sigma}(x)^4\right|\\
\rightarrow \frac{|C_\text{new}(x)-1|}{|C(x)-1|} &\approx \frac{12f''(x)^2\tilde{\sigma}(x)^2}{f'(x)^4}.\label{eqn:iterated_closure_ratio_app}
\end{align}
If the ratio in Eq.~\ref{eqn:iterated_closure_ratio_app} is greater than 1, then the closure gets worse after a second iteration of numerical inversion. $\hspace{1 cm} \Box$

\section{Corrected Numerical Inversion Calculation}
\label{sec:corrected_numerical_inversion_calculation}
With $Y\mapsto Z_\text{corr} = g^{-1}(Y)$, we will get a corrected calibrated distribution $\rho_{Z_\text{corr}|X}(z|x)$. For brevity, let $\rho_{Z_\text{corr}}(z)=\rho_{Z_\text{corr}|X}(z|x)$, where it is understood we are examining the distributions around a particular value of $x$. We will again require that $g'(x)>0$, so that
\begin{align}
\rho_{Z_\text{corr}}(z) = g'(z)\rho_Y(g(z)).
\end{align}
The closure condition is then equivalent to the condition
\begin{align}
\rho'_{Z_\text{corr}}(x) = 0,
\end{align}
i.e., the mode of the distribution $Z_\text{corr}|X=x$ occurs at $x$.  We have that
\begin{align}
\rho_{Z_\text{corr}}'(z) = g''(z)\rho_Y(g(z))+g'(z)^2\rho'_Y(g(z)),
\end{align}
so that the closure condition requires
\begin{align}
0 &=\rho_{Z_\text{corr}}'(x)\nonumber\\
&=g''(x)\rho_Y(g(x))+g'(x)^2\rho'_Y(g(x))\nonumber\\
\rightarrow 0 &=g''(x)+g'(x)^2\frac{\rho'_Y(g(x))}{\rho_Y(g(x))}.
\end{align}
We suppose that $g(x)$ is close to $f(x)$, $g(x)=f(x)+\alpha(x)$, with $|\alpha(x)|\ll \tilde{\sigma}(x)$. Then we have directly from the supposition that the distribution $Y|X=x$ is approximately Gaussian about its mode $f(x)$ with width $\tilde{\sigma}(x)$ that
\begin{align}
\frac{\rho'_Y(g(x))}{\rho_Y(g(x))} &= -\frac{\left(g(x)-f(x)\right)}{\tilde{\sigma}(x)^2}.
\end{align}
Then, the closure condition gives
\begin{align}
0 &=g''(x)+g'(x)^2\frac{\rho'_Y(g(x))}{\rho_Y(g(x))}\nonumber\\
&=g''(x)-g'(x)^2\frac{g(x)-f(x)}{\tilde{\sigma}(x)^2}.
\end{align}

\section{Corrected Numerical Inversion Parameterization}
\label{sec:corrected_numerical_inversion_parameterization}
We parameterize the corrected calibration function $g(x) = g(x;f(x);a_1,...,a_n)$. For the toy model used in this note, we use the parameterization
\begin{align}
g(x) = f(x)+\frac{a_1}{1+\exp(\frac{x-a_2}{a_3})}.
\label{eqn:app_parameterization}
\end{align}

In the model considered here, and for the response functions at the LHC, the closure goes to $1$ for large $x$ and moves away from $1$ for small $x$, a natural result of Eq.~\ref{eqn:closure_mode_text}. Thus, the parameterization in Eq.~\ref{eqn:app_parameterization} includes a ``turn-off'' to recover $g(x)=f(x)$ at large $x$ (with $a_3>0$).

In practice, there is some smallest value $x=x'$ which is being studied, and which per the discussion in the above paragraph tends to have the largest non-closure. The value $x'=20$ GeV is used in this note, which is the lowest calibrated $E_\text{T}$ at current conditions at the LHC. For the corrected calibration curve shown in Figure~\ref{fig:mode_closure_bigs}, the parameters $a_1,a_2,a_3$ are scanned over to minimize the non-closure at this value $x'$. For the corrected calibration curve shown in Figure~\ref{fig:mode_closure_bigs}, the values $a_2=a_3=x'=20$ GeV and $a_1 = 5$ GeV were used.

\clearpage
\newpage
\section{Toy Model of the ATLAS/CMS Response Function}
\label{sec:toy_model}
All the ``Proofs'' quoted in this note are valid in general, regardless of the response function $R(x)$ and the underlying distributions $Y|X=x$ (within the assumptions outlined in Section~\ref{sec:assumptions}). We also expect that the ``Derivations'', which are all approximate formulas, to apply in a wide variety of cases. In order to visualize some of the results, and verify the approximations, a particular model was needed in order to get numerical values. All figures made in this note were derived from a simple model of the ATLAS or CMS jet $E_\text{T}$ response function\footnote{Energies are measured with calorimeters and momenta are measured with tracking detectors.  In-situ corrections using momentum balance techniques constrain the momentum.  For small-radius QCD jets, the $E_\text{T}$ and $p_\text{T}$ are nearly identical.  Since the simulation-based correction of calorimeter jets is used here as a model, the $E_\text{T}$ is used throughout. }. After specifying $f(x)$ and the distributions $Y|X=x$, the calibrated distributions were constructed using the analytic form of the calibrated distributions Eq.~\ref{eqn:newdist}. Then the various moments were found numerically for the calibrated distribution at each value $x$.

The response function was guided both by physical intuition and by the intention to reasonably simulate response functions published by ATLAS~\citep{Aad:2011he} and CMS~\citep{Chatrchyan:2011ds,Khachatryan:2016kdb}. When there is only a small amount of energy already in a detector cell, the detector only reconstructs a small fraction of the energy put into it, because of noise thresholds and the non-compensating nature of the ATLAS and CMS detectors. Whereas if there is already a lot of energy in a detector cell, the detector reconstructs almost all of the energy put into it. Thus $f'(x)$ was designed  to be low at low values of $x$ and then to rise steadily to 1 at high values of $x$.  This intuition does not directly apply to jets that directly use tracking information (e.g. particle-flow jets in CMS), but the for the sake of simplicity only one (calorimeter) jet definition is used for illustration.

\begin{figure}[h!]
\begin{center}
\includegraphics[width=0.9\textwidth]{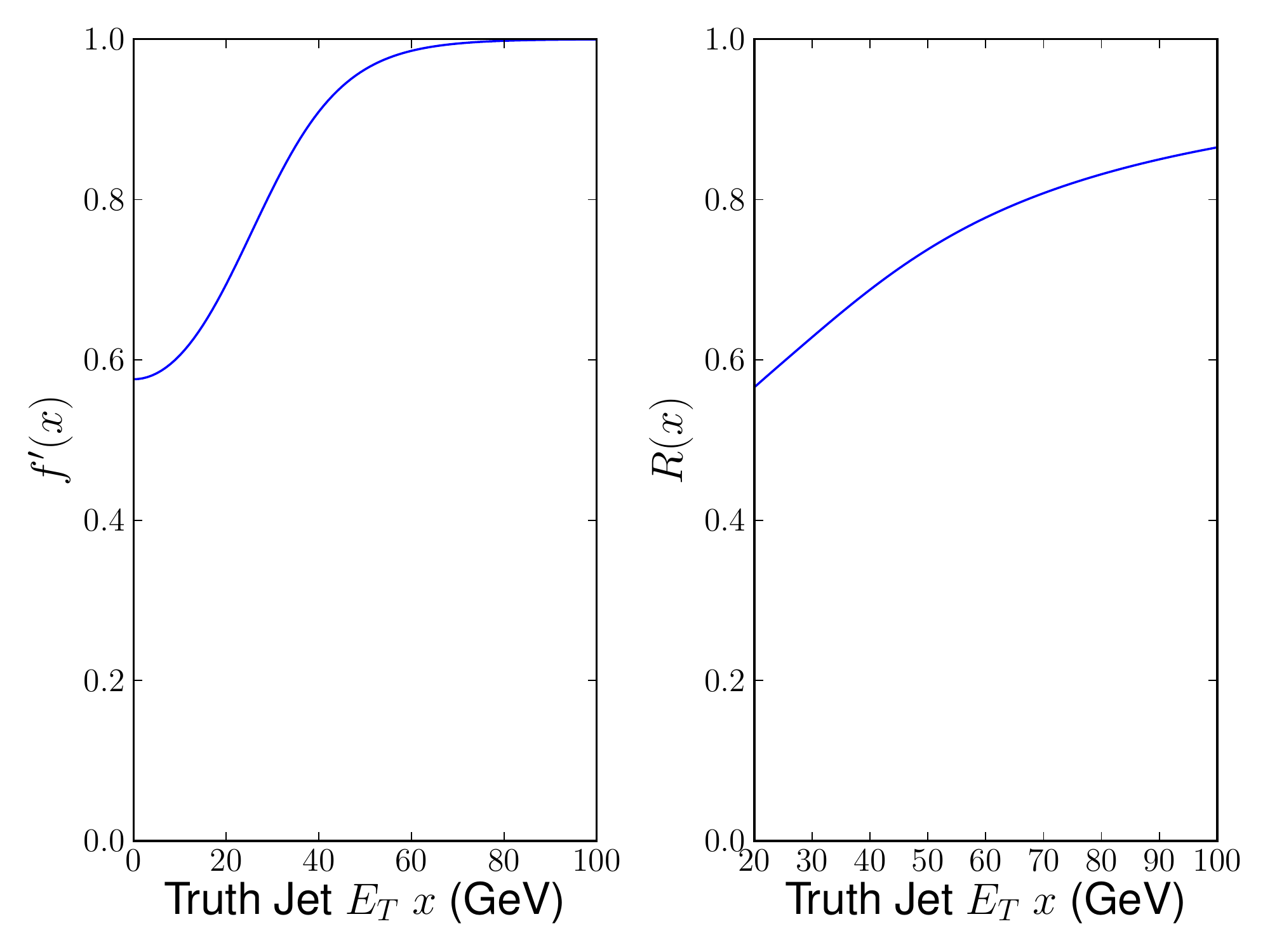}
\end{center}
\caption{The toy model used in this note to simulate conditions in ATLAS or CMS. The left plot shows $f'(x)$ and the right plot shows $R(x)$.}
\label{fig:model}
\end{figure}

$f'(x)$ was the integrated to get $f(x)$ and divided by $x$ to get $R(x)$. The resulting $R(x)$ distribution approximately corresponds to the $R=0.4$ anti-$k_t$~\citep{Cacciari:2008gp} central jet response at the EM scale available in Ref.~\citep{Aad:2011he} (e.g. Fig. 4a). The shapes of $f'(x)$ and $R(x)$ in this model can be seen in Figure~\ref{fig:model}.

In this simplified model, the distributions $Y|X=x\sim\mathcal{N}(f(x),\sigma(x))$ were used. In ATLAS and CMS, $Y|X=x$ is approximately Gaussian. The constant value of $\sigma(x)=7$ GeV was used, corresponding to a calibrated resolution (Fig.~\ref{fig:mean_resolution}) of about 50\% at $E_\text{T}=20$ GeV.  This is consistent with e.g. Ref.~\citep{Aad:2015ina} and has the property that $\sigma'(x) = 0$, which should be the case if pileup is the dominant contributor to the resolution of low $E_\text{T}$ jets.
\clearpage
\newpage

\bibliographystyle{elsarticle-num}
\bibliography{myrefs.bib}{}

\begin{thebibliography}{10}

\bibitem{Bhatti:2005ai}
{{\bf CDF} Collaboration}, ``{Determination of the jet energy scale at the
  collider detector at Fermilab},'' {\em Nucl. Instrum. Meth.}, vol.~A566,
  pp.~375--412, 2006.

\bibitem{Abazov:2013hda}
{{\bf D0} Collaboration}, ``{Jet energy scale determination in the D0
  experiment},'' {\em Nucl. Instrum. Meth.}, vol.~A763, pp.~442--475, 2014.

\bibitem{Cacciari:2008gp}
M.~Cacciari, G.~P. Salam, and G.~Soyez, ``{The anti-$k_t$ jet clustering
  algorithm},'' {\em JHEP}, vol.~0804, p.~063, 2008.

\bibitem{topo1}
{W. Lampl et al.}, ``{Calorimeter clustering algorithms: description and
  performance},'' no.~ATL-LARG-PUB-2008-002, 2008.

\bibitem{topo2}
{C. Cojocaru et al.}, ``{Hadronic calibration of the ATLAS liquid argon end-cap
  calorimeter in the pseudorapidity region $1.6 < |\eta| < 1.8$ in beam
  tests},'' {\em Nucl. Instrum. Meth.}, vol.~A531, p.~481, 2004.

\bibitem{pflow1}
{{\bf CMS} Collaboration}, ``{Particle Flow Event Reconstruction in CMS and
  Performance for Jets, Taus, and $E_{T}^\text{miss}$},''
  no.~CMS-PAS-PFT-09-001.

\bibitem{pflow2}
{{\bf CMS} Collaboration}, ``{Commissioning of the Particle-Flow Reconstruction
  in Minimum-Bias and Jet Events from $pp$ Collisions at 7 TeV},''
  no.~CMS-PAS-PFT-10-002.

\bibitem{Aad:2011he}
{{\bf ATLAS} Collaboration}, ``{Jet energy measurement with the ATLAS detector
  in proton-proton collisions at $\sqrt{s}=7$ TeV},'' {\em Eur. Phys. J.},
  vol.~C73, no.~3, p.~2304, 2013.

\bibitem{Chatrchyan:2011ds}
{{\bf CMS} Collaboration}, ``{Determination of Jet Energy Calibration and
  Transverse Momentum Resolution in CMS},'' {\em JINST}, vol.~6, p.~P11002,
  2011.

\bibitem{Khachatryan:2016kdb}
``{Jet energy scale and resolution in the CMS experiment in pp collisions at 8
  TeV},'' {\em Submitted to: JINST}, 2016.

\bibitem{JetEtmissApproved2013HighMuPileup}
``{Pileup effects on jet calibration and response at high $\mu$},'' {\em
  JetEtmissApproved2013HighMuPileup}, 2013.

\end{thebibliography}

\end{document}